\documentclass{aastex}
\usepackage{emulateapj5}
\usepackage{apjfonts}

\slugcomment{ApJ accepted version}

\shorttitle{Shapes and Positions of Black Hole Shadows in Accretion Disks and Spin
Parameters of Black Holes}
\shortauthors{Rohta Takahashi}

\begin{document}

\title{Shapes and Positions of Black Hole Shadows in Accretion Disks \\
and Spin Parameters of Black Holes}

\author{Rohta Takahashi\altaffilmark{1}}
\affil{Yukawa Institute for Theoretical Physics,
Kyoto University, Sakyo-ku,  Kyoto 606-8502, Japan}

\altaffiltext{1}{e-mail: rohta@yukawa.kyoto-u.ac.jp}

\begin{abstract}
Can we determine a spin parameter of a black hole 
by observation of a black hole shadow in an accretion disk? 
In order to answer this question, 
we make a qualitative analysis and a quantitative analysis 
of a shape and a position of a black hole shadow 
casted by a rotating black hole on an optically thick accretion disk and its dependence 
on an angular momentum of a black hole. 
We consider two types of inner edges, an event horizon and a marginally stable orbit,  
of accretion disks. 
We have found black hole shadows with a quite similar size and a shape for largely 
different black hole spin parameters and a same black hole mass. 
Thus, it is practically difficult to determine a spin parameter of a black hole 
from a size and a shape of a black hole shadow in an accretion disk.  
On the other hand, 
a displacement of a black hole shadow from a rotation axis of a black hole is caused by 
a frame dragging effect in the vicinity of a black hole. 
The extent of the displacement largely depend on a black hole spin parameter. 
However, it is difficult to determine a position of a rotation axis of a black hole 
observationally. 
So, we can not practically measure a spin parameter of a black hole from the displacement from 
a rotation axis of a black hole shadow. 
We newly introduce a bisector axis of a black hole shadow named {\it a shadow axis}. 
We define this shadow axis as a bisector 
perpendicular to a line segment of the maximum width of a black hole shadow. 
We can determine a position of a shadow axis of  
by observation of a black hole shadows. 
For a non-rotating black hole,  
the minimum interval between a mass center of a black hole 
and a shadow axis is null. 
On the other hand, for a rotating black hole  
a shape and a position of a black hole shadow are not symmetric 
with respect to a rotation axis of a black hole shadow. 
So, in this case the minimum interval between 
a mass center of a black hole and a shadow axis is finite. 
An extent of this minimum interval is roughly proportional to 
a spin parameter of a black hole for a fixed inclination angle between 
a rotation axis of a black hole and a direction of an observer. 
The maximum extent of these minimum intervals is about $1.5 r_g$.  
Here, $r_g$ is a gravitational radius. 
This is realized when the maximum inclination angle and the maximally rotating black hole in 
an accretion disk with its inner edge of an event horizon. 
In order to measure a spin parameter of a black hole, 
if a shadow axis is determined observationally, 
it is crucially important to determine 
a position of a mass center of a black hole in a region of a black hole shadow. 
We also discuss how to determine a mass center of a black hole by observation of 
a black hole shadow in an accretion disk. 
\end{abstract}


\section{INTRODUCTION}
During the last century, masses of supermassive black holes and stellar 
mass black holes are observationally measured. 
Determination of a spin parameter of a black hole is one of the
greatest challenges in astrophysics in this century. 
So far, several determination methods of a black hole spin are proposed. 
Broad and/or skewed X-ray Fe K$\alpha$ emission lines, first discovered by ASCA,  
are extensively studied 
(e.g. Fabian et al. 1995, Tanaka et al. 1995, Bromley, Miller \& Pariev 1998).  
If appropriate accretion disk models in Kerr spacetime are assumed, 
one can explain these broadening features for some spin parameters 
of central black holes. 
Some authors actually propose some values of spin parameters 
of supermassive black holes (e.g. Miller et al. 2002). 
Melia et al. (2001) said that they constrain the black hole spin in 
Sagittarius A* by sub-millimeter timing observations. 
Recently, by high-resolution infrared observations of Sagittarius A*
Genzel et al. (2003) estimates the spin parameter of the massive 
black hole in Sgr A*. 
These two methods crucially depend on a period of a matter rounding 
marginally stable orbit with Keplerian velocity. 
In the case of a optically thin accretion flow like Bondi-type flow 
in which 
an angular momentum of an accreting matter is smaller than Keplerian and 
radial velocity of the matter is large, 
it is open to question 
whether a period of a marginally stable orbit determine a minimum time 
of variability. 
Other methods to determine a spin parameter of a black hole are proposed.  
X-ray polarized emission from an accretion disk around a black hole  
(e.g. Conners, Stark \& Piran 1980), 
microlensing light curves due to a black hole as a lensing object 
(e.g. Asada, Kasai \& Yamamoto 2003)
and  
gravitational wave emission from a star orbiting around a black hole 
(e.g. Tanaka et al. 1996, Finn \& Thorne 2000) 
are also influenced by a black hole rotation.    

Observation of a black hole shadow is 
another possible method to determine a spin parameter of a rotating black hole 
in future. 
This is because in the vicinity of a black hole null geodesics directly 
reflect effects of black hole rotation, such as frame dragging. 
As Falcke, Melia \& Agol (2000) pointed out, in the case of Sgr A*  
radio interferometers with sufficient spatial resolutions may 
resolve the central black hole shadows in the optically thin emitting 
medium by observations of sub-millimeter wavelength. 
A shape of a black hole shadow in an optically thin medium are 
deformed by black hole rotation 
(e.g. Bardeen 1973, Chandrasekhar 1983, Falcke, Melia \& Agol 2000). 
An apparent shape of such black hole shadow in the $(x,~y)$-plane 
is described by (Bardeen 1973, Chandrasekhar 1983) 
\begin{eqnarray}
x&=&\frac{1}{a\sin{i}}
	\left[
	t^2+a^2-3m^2-\frac{2m(m^2-a^2)}{t}
	\right],
\label{shadow0}\\
y^2&=&\frac{(m+t)^3}{a^2}
	\left[
	-t+3m-\frac{4m(m^2-a^2)}{t^2}
	\right] \nonumber \\
	&&+(a^2-x^2)\cos^2{i}, 
\label{shadow1}
\end{eqnarray}
where $m$ and $a$ are a mass of a black hole and an angular momentum 
per unit mass of a black hole, respectively, 
$i$ is an inclination angle between a rotation axis 
of a black hole and a direction to an observer and $t$ is an intervening parameter. 
Here, $x$ and $y$ are normalized by a gravitational radius $r_g$ 
defined as $r_g=Gm/c^2$, 
where $G$ is the gravitational constant and $c$ is the speed of light. 
In the case of large inclination $i\sim 90^\circ$ and a large spin parameter 
$a/m\sim 1$, 
a black hole shadow have an elongated shape in a direction of a rotation axis of a 
black hole (e.g. Bardeen 1973, Chandrasekhar 1983, Falcke, Melia \& Agol 2000). 
If such an elongation feature of a black hole shadow in optically thin emitting 
medium is observed, the spin parameter of the black hole can be determined 
by fitting a contour of a black hole shadow given by equations (\ref{shadow0}) 
and (\ref{shadow1}) to the observed image. 

On the other hand, in the case of a black hole in an optically thick and 
geometrically thin accretion disk, 
is it possible to measure a spin parameter of a black hole 
from observation of a black hole shadow? 
One of the main purposes of the present paper is to answer this. 
A contour of a black hole shadow in an accretion disk is basically calculated 
by Cunningham \& Bardeen (1972) and Cunningham (1973, 1975) for a rotating 
black hole. 
Luminet (1979) and Fukue \& Yokoyama (1988) calculate apparent images of 
black hole shadows in accretion disks for non-rotating black holes. 
In the case of a rotating black hole, 
such as a deformed black hole shadow in optically thin medium as mentioned above, 
one might firstly think that a spin parameter of a black hole can be determined 
from a deformed shape of a black hole shadow due to a black hole rotation. 
Actually, many authors calculate such deformed shadows due to black hole 
rotation in various contexts 
(Cunningham \& Bardeen 1972, Cunningham 1973, 1975, Sikora 1979, 
Perez \& Wagoner 1991, Jaroszynski, Wambsganss, \& Paczy\'{n}ski 1992, Kindl 1995, 
Quien, Wehrse \& Kindl 1996, Begelman \& Rees 1996,  
Hollywood \& Melia 1995, 1997, Bromley, Miller \& Pariev 1998, Pariev \& Bromley 1998, 
Bromley, Melia \& Liu 2001).  
In the present paper, we show black hole shadows in accretion disks 
with quite similar shapes for largely different spin parameters of central black holes. 
When we calculate a shape of a black hole shadow in an accretion disk, 
an innermost radius of a luminous part of an accretion disk is significantly important. 
Traditionally, as an emitting disk surrounding a black hole, 
the standard disk (Shakura \& Sunyaev 1973, Novikov \& Thorne 1973) is used. 
In such a disk, the innermost radius of the luminous part of the accretion disk 
corresponds to a marginally stable orbit if we neglect dimming effects 
due to gravitational redshift in the vicinity of the black hole. 
Since the surface density inside this edge is expected to 
be extremely smaller than that outside the edge, 
it is usually assumed that there is no luminous matter inside the inner edge. 
However, it has not been proven whether the region inside the inner edge 
is non-luminous or not (Fukue 2003). 
Reynolds and Begelman (1997) discuss in detail the possibility that the material 
within innermost stable orbit may be bright in the X-ray iron line. 
Recently, 
from relativistic magnetohydrodynamical simulations, 
Krolik \& Hawley (2002) proposed that the inner edge of an accretion disk 
around a black hole depends on the definition, i.e. turbulence edge, 
stress edge, reflection edge, and radiation edge are defined by each physical process. 
In the present study, we define the inner edge as an innermost radius of 
a luminous part of an accretion disk.  
Also, in the case of a super-Eddington accretion disk, 
an optically-thick luminous disk extends from a large radial distance 
to an event horizon (e.g. Watarai \& Mineshige 2003), 
i.e. 
an innermost radius of a super-Eddington accretion disk 
is smaller than a marginally stable orbit. 
This is because an effective potential based on an angular momentum distribution 
does not have a potential minimum for accretion rates, 
$\dot{M}>10L_{\rm E}/c^2$, where $L_{\rm E}$ is the Eddington luminosity 
(Watarai \& Mineshige 2003). 

In the present study, we calculate contours of black hole shadows 
in geometrically thin and optically thick accretion disks for two kinds of innermost radii, 
a marginally stable orbit and an event horizon. 
When an innermost radius of an accretion disk is not close to an event 
horizon, the radiation from the underside of the accretion disks 
appear within a black hole shadow consisting of the direct orbits. 
Since the intensity of such radiation is not strong, we neglect such radiation. 
In \S 2, calculation methods are shown. 
We give a qualitative analysis and a quantitavive analysis of the shape and position 
of the black hole shadow casted by the rotating black hole in the center of 
the accretion disk for two kinds of innermost radii of accretion
disks in \S 3.  
We present quite similar black hole shadows for largely different 
spin parameters of black holes in \S4. 
We give discussion and conclusions in \S5. 

\section{Calculation Methods of Black Hole Shadows in Accretion Disks}
In this section, we briefly show a calculation method of a contour 
of a black hole shadow in Kerr spacetime. 
The propagation of radiation from a matter around a Kerr black hole 
is extensively studied by Cunningham (1975). 
The background metric is written by the Boyer-Lindquist coordinates with 
$c=G=1$, 
\begin{eqnarray}
ds^2&=&\left(1-\frac{2mr}{\Sigma}\right)dt^2
+\frac{4amr\sin^2\theta}{\Sigma}dtd\phi \nonumber\\
&&-\left(r^2+a^2+\frac{2a^2mr\sin^2\theta}{\Sigma}\right)\sin^2\theta 
d\phi^2
\nonumber \\
&&-\frac{\Sigma}{\Delta}dr^2-\Sigma d\theta^2,
\end{eqnarray}
where $\Delta\equiv r^2-2mr+a^2$, $\Sigma\equiv r^2+a^2\cos^2\theta$, and 
$m$ and $a$ denote the mass and angular momentum per unit mass of the black 
hole, respectively. 
In the present study, we assume an axisymmetric emitting matter distribution 
with respect to a rotation axis of a black hole. 
In such a case, a rotation axis of a black hole is equivalent to a rotation axis of an accretion disk. 
In the present study, we only consider a co-rotating accretion disk around a rotating black hole. 
Since we assume the axisymmetric matter distribution with respect to 
a rotation axis of a black hole, 
we only need to calculate projected null geodesics on the $(r,~\theta)$ plane 
in order to calculate a contour of a black hole shadow. 
An integral form of the geodesic equations of $(r,~\theta)$-component 
can be written as 
(e.g. Carter 1968, Misner, Thorne \& Wheeler 1973, Chandrasekhar 1983)
\begin{equation}
\int^\theta_{\theta_0}\frac{d\theta}{[\Theta(\theta)]^{1/2}}
=\pm\int^r_{r_0}\frac{dr}{[R(r)]^{1/2}}, 
\label{rtintegral}
\end{equation}
where 
\begin{eqnarray}
\Theta(\theta)&=&q^2+a^2\cos^2\theta-p^2\cot^2\theta,\\
R(r)&=&r^4+(a^2-p^2-q^2)r^2+2m[q^2+(a-p)^2]r-q^2a^2. 
\end{eqnarray}
Here, $p$ and $q$ are constants along the geodesic and defined as 
\begin{equation}
p=L_z/E,~~~q^2=C/E^2, 
\end{equation}
where $E$, $L_z$ and $C$ are an energy, an angular momentum and Carter's 
constant along the geodesic, respectively. 
Two impact parameters $(x,~y)$ in the plane of the sky of an observer 
can be expressed by 
the two constant, $p$ and $q$, as
(Bardeen 1973, Cunningham \& Bardeen 1972, Chandrasekhar 1983)
\begin{eqnarray}
x&=&-p/\sin{i},\\
y&=&q^2+\cos^2{i}-p^2\cot^2{i},  
\end{eqnarray}  
where $i$ is an inclination angle between a rotation axis of 
accretion disk and a direction to an observer. 
As several authors have done, 
$\theta$ and $r$ integrals of 
geodesics can be solved analytically by using Elliptic integrals 
of the first kind and the second kind 
(e.g. Cunningham 1975, Rauch \& Blandford 1994). 
These analytic solutions have many different sub-cases depending on the values of 
constants along the trajectory and the spin parameter of the black hole. 
Also, the signs of the equation (\ref{rtintegral}) are chosen independently 
(Carter 1968) and depend on the direction of the geodesics.  
Rauch \& Blandford (1994) extensively studied these analytic solutions and the calculation method 
of the geodesics in a Kerr metric. 
All patterns of analytic solutions of equation (\ref{rtintegral}) 
are listed in tables in Appendix A of Rauch \& Blandford (1994). 
Note that 
while Rauch \& Blandford (1994) give all analytic solutions of $r$-$\theta$ components of the 
geodesics in a Kerr metric, 
Rauch \& Blandford (1994) do not present the analytic solutions of 
the $\phi$-component and the $t$-component of the geodesics.
For details of the calculation methods of the geodesics in a Kerr metric, 
refer to Rauch \& Blandford (1994). 
In the present study, we use the analytic formulas of equation (\ref{rtintegral}).   
If we consider an observer at infinity from a black hole, 
in the equation (\ref{rtintegral}) we set $r$ and $\theta$ 
components of 
an observer to be $(r_0,~\theta_0)=(\infty,~i)$. 
When we need to specify $r_0$, we use a value $r_0=10^6$ 
(in units of $r_{\rm g}$). 
In the subsequent sections we calculate black hole shadows in 
thin accretion disks 
and we set a $\theta$ component of an end point of a geodesics to 
be $\theta=\pi/2$ in the equation (\ref{rtintegral}). 
In terms of an innermost radius of a luminous part of an accretion disk, 
we consider two types of radii as an example; 
an event horizon $r_{{\rm h}+}$ and a marginally stable orbit $r_{\rm ms}$. 
An event horizon of a black hole defined as 
$r_{{\rm h}+}=m+(m^2-a^2)^{1/2}$, and an explicit form of a marginally 
stable orbit $r_{\rm ms}$ can be seen in Bardeen, Press \& Teukolsky (1972). 
Scales of black hole shadows in the present study are normalized by a gravitational 
radius, $r_g$, defined as $r_g=Gm/c^2$. 

\section{Black Hole Shadows in Geometrically Thin Accretion Disks}


\subsection{Qualitative Features}

In order to understand how a contour of a black hole shadow depend on 
a spin parameter, $a/m$, of a black hole, an inner edge, $r_{\rm in}$, 
of an accretion disk and 
an inclination angles, $i$,  
we plot in figure 1 
contours of black holes shadows in optically thick accretion disks 
for $a/m=0$, $0.5$ and $0.998$ 
(from {\it left panels} to {\it right panels}), 
$r_{\rm in}=r_{{\rm h}+}$ ({\it thick solid lines}) 
and $r_{\rm ms}$ ({\it thin solid lines}), 
and $i=0^\circ$, $45^\circ$ 
$80^\circ$ and $89^\circ$ (from {\it top panels} to {\it bottom panels}). 
We also plot  
contours of black hole shadows casted by black holes 
in optically thin emitting matters 
(e.g. Bardeen 1973, Chandrasekhar 1983, Falcke, Melia \& Agol 2000) 
by dashed lines in figure 1. 
These shadows are calculated from equations (\ref{shadow0}) and (\ref{shadow1}).  
As a reference frame, in all panels of figure 1, 
we also plot lines of $x=0$ and those of $y=0$ by dotted lines. 
The origin, $(x,~y)=(0,~0)$, is a position of a black hole, i.e. a mass center of a black hole. 
In this subsection we summarize qualitative features of contours of black hole shadows 
in accretion disks. 
These features are partly investigated in references cited in \S1. 
In subsequent subsections, we investigate black hole shadows quantitatively.  

First, we see how a contour of a black hole shadow depends on an inclination angle, $i$. 
In the case of $i=0^\circ$, a shape of a black hole shadow is circle with its  
center at the origin, i.e. a position of a black hole mass center, 
for any black hole spin parameters because of axisymmetric luminous 
matter distribution with respect to a rotation axis of a black hole.  
Although a size of a black hole shadow for an optically thin medium plotted by a dashed line 
is nearly independent of a black hole spin parameter, 
for both cases of shadows in disks with 
$r_{\rm in}=r_{{\rm h}+}$ (thick solid lines) and $r_{\rm sm}$ (thin solid lines) 
the size of the black hole shadow become smaller for a more rapidly rotating black hole. 
This is simply because a radius of 
an event horizon and a marginally stable orbit become smaller 
for a more rapidly rotating black hole. 
In the case of finite inclination angles, i.e. $i=45^\circ$, $80^\circ$ and $89^\circ$ in figure 1, 
since accretion disks non-axisymmetrically prevent null geodesics 
from plunging into a black hole horizon, 
black hole shadows for cases $r_{\rm in}=r_{{\rm h}+}$ and $r_{\rm sm}$ 
deviate from circles, and slightly become flat shapes. 
In the case of a black hole shadow for optically thin medium, 
since there is no occultation by an accretion disk 
the shape of the black hole shadow is nearly circle 
even for a finite inclination angle and a finite value of a black hole spin parameter. 
Especially, in the case of $i=89^\circ$ the black hole shadow in an accretion disk 
for cases $r_{\rm in}=r_{{\rm h}+}$ or $r_{\rm sm}$, 
underside of a black hole shadow which is seen in the case of a shadow in optically thin medium 
is partly occulted by an accretion disk. 
When an outer radius of an accretion disk is not so large, a black hole shadow consists of 
radiation from a backside of an accretion disk can be seen 
when viewing from nearly edge-on, $i\sim 90^\circ$. 
In the present study, we assume sufficiently large accretion disks and do not consider 
radiation from the backside of an accretion disk. 

In the case of $r_{\rm in}=r_{{\rm h}+}$, a horizontal size of a black hole shadow becomes 
larger for a larger inclination angle. 
In the case of a finite inclination angle, 
in a region of an upper side, i.e. $y>0$, 
a length of a geodesics from an observer to a crossing point on an accretion disks 
becomes longer than a geodesics in the case of $i=0^\circ$. 
In such cases, some geodesics crossing accretion disks 
when $i=0^\circ$ directly plunge into black hole horizon, 
and a size of a black hole shadow becomes larger for a larger inclination angle. 
This effect is strong near a black hole. 
In the case of $r_{\rm in}=r_{\rm ms}$ 
a size of a black hole shadow becomes slightly larger for a larger inclination angle. 
But in the case of $r_{\rm in}=r_{\rm ms}$ this effect is weaker than cases 
when $r_{\rm in}=r_{{\rm h}+}$. 
This is because when a marginally stable orbit is larger than 
an event horizon for a fixed spin parameter and in the case of $r_{\rm in}=r_{{\rm h}+}$ 
geodesics consisting contours of black hole shadows are more strongly bent than those 
in the case of $r_{\rm in}=r_{\rm ms}$. 

Additionally, in the case of $r_{\rm in}=r_{\rm ms}$, a nearly maximum inclination angle, 
$i\sim 90^\circ$, and a non-rotating black hole, as seen in the left bottom panel in figure 1, 
a black hole shadow has two tips near $y\sim0$ and $x\sim\pm7$. 
Upper side of a black hole shadow is made by geodesics from a back of a black hole and 
these geodesics are strongly bent by a strong gravitational field of a black hole. 
In the region relatively apart from a black hole, these bending effects are not so strong 
and geodesics become nearly Newtonian. 
In such cases, black hole shadows have tips which reflect nearly Newtonian geodesics. 
So, if we put an accretion disk with an inner edge larger than a marginally stable orbit 
and we view such shadows with clear tips when a large inclination angle. 

As mentioned above, in the case of a rotating black hole 
for a larger spin parameter of a black hole 
an event horizon and a marginally stable orbit become smaller and a size of a black hole 
shadows also become smaller. 
But, this smallness of a black hole shadow is not unique properties of 
effects of black hole rotation. 
For a constant spin parameter of a black hole, a size of a black hole shadow becomes smaller 
for a smaller inner edge of an accretion disk. 
Also, for a constant spin parameter of a black hole and a constant inner edge of an accretion 
disk, a size of a black hole shadow becomes smaller for a smaller black hole mass. 
What is the unique property of black hole rotation? 
Although in the case of a non-rotating black hole a black hole shadow is perfectly symmetric 
with respect to a rotation axis of an accretion disk, 
as shown in figure 1, in the case of a rotating black hole 
a black hole shadow becomes non-axisymmetric with respect to a rotation axis. 
This is because geodesics near a black hole suffers from effects of frame dragging and 
such a geodesic becomes non-axisymmetric with respect to a rotation axis.  
This non-axisymmetric shape of the black hole shadow can be also seen in 
past studies 
(Cunningham \& Bardeen 1972, Cunningham 1973, 1975, Sikora 1979, 
Perez \& Wagoner 1991, Jaroszynski, Wambsganss, \& Paczy\'{n}ski 1992, Kindl 1995, 
Quien, Wehrse \& Kindl 1996, Begelman \& Rees 1996,  
Hollywood \& Melia 1997, Bromley, Miller \& Pariev 1998, Pariev \& Bromley 1998, 
Bromley, Melia \& Liu 2001).  
If luminous matters around a black hole are distributed non-axisymmetrically, 
the non-axisymmetric property of a black hole shadow is not unique effect of 
black hole rotation. 


\subsection{Maximum Size of a Black Hole Shadow}
In this section, we see quantitative features of the maximum size 
of a black hole shadow. 
In figure 2, 
we show dependence of the maximum size of a black hole shadow 
on a spin parameter of a black hole and an inclination angle  
for three kinds of surrounding mediums; 
an accretion disk 
with innermost radius of $r_{\rm in}=r_{{\rm h}+}$ ({\it dotted lines})
and an accretion disk with $r_{\rm in}=r_{\rm ms}$ ({\it solid lines}), 
and optically thin emitting medium ({\rm dash-dotted lines}). 
Note that the maximum size of a black hole shadow generally does not coincide 
with the maximum horizontal size of a black hole shadow 
because of black hole rotation. 
In figure 1, 
this feature can be clearly seen in black hole shadows in disks  
for $a/m=0.998$ and $i=80^\circ$. 

We can see gravitational focusing effects of geodesics in figure 2. 
For example, in the case of a non-rotating black hole 
a radius of a marginally stable orbit is 6. So, a diameter of a circle of a 
marginally stable orbit is 12. 
However, for an inclination angle $i=0^\circ$, $r_{\rm in}=r_{\rm ms}$ 
and a non-rotating black hole the maximum size of a black hole shadow is nealy 13.85 
in figure 2 because of gravitational focusing effects. 

For any spin parameter of a black hole, 
the maximum size of a black hole shadow becomes larger 
for a larger inclination angle. 
The reason for this is explained in the previous section. 
For any spin parameter of a black hole and a fixed inclination angle 
a black hole shadow in an accretion disk with $r_{\rm in}=r_{{\rm h}+}$ is smallest 
among black hole shadows for these three kinds of surrounding medium. 
In the case of a slowly rotating black hole or a non-rotating black hole 
($a/m\lesssim 0.5$), 
the maximum size of a black hole shadow around optically thin medium 
is 
among those in the cases of the other two disks for same inclination angle.  
On the other hand, in the case of a rapidly rotating black hole ($a/m\gtrsim 0.8$) 
the maximum size of a black hole shadow around optically thin medium 
is larger than the black hole shadows in other two kinds of accretion disks 
for a fixed inclination angle. 
Although the maximum size of a black hole shadow in optically thin medium 
concentrates on a narrow range of a size 
(i.e. from $\sim 9.6$ to $\sim 10.4$), 
the maximum size of a black hole shadow in an accretion disk expands 
a relatively wide range of a size. 
This is because the event horizon and marginally stable orbit 
depend on a spin parameter of a black hole as mentioned above. 

In the case of an accretion disk with $r_{\rm in}=r_{{\rm h}+}$, 
the maximum size of a black hole shadow is $\sim 8.7$ for 
$i=90^\circ$ and $a/m=0$. 
On the other hand, the minimum size of a black hole shadow is 
$\sim 4.0$ for $i=0^\circ$ and $a/m=1$. 
When an accretion disk with $r_{\rm in}=r_{\rm ms}$, 
the maximum size of a black hole shadow is $\sim 14.7$ for 
$i=90^\circ$ and $a/m=0$, and  
the minimum size of a black hole shadow is $\sim 4.2$ 
for $i=0^\circ$ and $a/m=1$. 
In contrast to the case of a black hole shadow 
in optically thin medium for which 
a size of a black hole shadow is in a narrow range as mentioned above, 
if an inner edge of an accretion disk is equal to or smaller 
than a marginally stable orbit, 
a size of a black hole shadow expands in a wider range from $4.0$ to 
$14.7$ than a black hole shadow in optically thin medium as shown in figure 2.  
This is a remarkable quantitative feature of a black hole shadow 
in an accretion disk. 

\subsection{Flatness of Black Hole Shadow}
Here, for a single black hole shadow we consider a ratio of the minimum width of 
a black hole shadow to the maximum width of a black hole shadow. 
This ratio is important because 
this ratio is a good indicator of an inclination angle 
between a rotation axis of a black hole and a direction of an observer. 
This ratio is similar to a flatness of a black hole shadow. 
This ratio is a dimensionless quantity. 
Note that a line segment of the minimum width and a line segment of the maximum width 
generally do not cross perpendicularly. 
In figure 3, 
we show dependence of the ratio on a spin parameter of a black hole   
for three inclination angles ($45^\circ$, $80^\circ$ and $89^\circ$) and 
two kinds of accretion disks with $r_{\rm in}=r_{{\rm h}+}$ 
({\it dotted lines}) 
and $r_{\rm in}=r_{\rm ms}$ ({\it solid lines}).  
In the case of an inclination angle of $i=0^\circ$, this ratio is unity. 
Although the ratio strongly depends on an inclination angle, 
the ratio weakly depends on a spin parameter of a black hole. 
In the both cases of black hole shadows in accretion disks with 
$r_{\rm in}=r_{{\rm h}+}$ and $r_{\rm in}=r_{\rm ms}$, 
for an inclination angle $i=89^\circ$ this ratio is nearly 0.5 for most of 
a range of a spin parameter. 
This value of $\sim0.5$ is a lower limit of the ratio. 


\subsection{Horizontal Shift of a Black Hole Shadow from a Rotation Axis 
of a Black Hole}
As mentioned in \S 3.1, 
if we consider an axisymmetric luminous matter distribution with respect to 
a rotation axis of a black hole,  
a horizontal shift of a black hole shadow, which is a displacement of a shadow center 
in the direction of $x$ axis, shows a significant unique property 
of effects of a black hole rotation.  
In order to quantify this shift, we define $x$ coordinates, $x_{\rm c}$, 
as $x_{\rm c}=(x_{\rm max}+x_{\rm min})/2$ where $x_{\rm max}$ and $x_{\rm min}$ 
are the maximum $x$-coordinate and the minimum $x$-coordinate of a contour 
of a black hole shadow, respectively. 
In the case of a non-rotating black hole, a shape and a position of a black hole shadow 
are symmetric with respect to a rotation axis of a black hole, i.e. $y$ axis. 
So, in this case, $x_c=0$.  
On the other hand, in the case of a rotating black hole  
a shape and a position of a black hole shadow is not symmetric 
with respect to a rotation axis of a black hole. 
So, in this case a value of $x_c$ is finite. 
This is because frame dragging effects of geodesics due to a black hole rotation 
break the symmetric property of a black hole shadow with respect to a rotation axis. 


In figure 4, we show 
the shift of the center of the black hole shadow in the direction of $x$ axis 
for three inclination angles ($45^\circ$, $80^\circ$ and $89^\circ$) 
and two kinds of 
accretion disks with $r_{\rm in}=r_{{\rm h}+}$ ({\it dotted lines}) 
and $r_{\rm in}=r_{\rm ms}$ ({\it solid lines}).  
In this figure, we can see that 
an extent of the horizontal shift is roughly proportional to 
a spin parameter of a black hole for a fixed inclination angle. 
An extent of the maximum horizontal shift is about $1.45$  
in the case of the maximum rotation, the maximum inclination angle and 
$r_{\rm in}=r_{{\rm h}+}$. 
A horizontal shift for the case of $r_{\rm in}=r_{{\rm h}+}$ is 
larger than 
that for the case of $r_{\rm in}=r_{\rm ms}$ for same inclination angle 
and same spin parameter of a black hole. 
This is because effects of frame dragging due to black hole rotation 
are more significant in a nearer region of a black hole. 
These strong frame dragging effects result in a large shift of 
a black hole shadow in $x$ direction. 
If we can measure this horizontal shift observationally, 
we can well constrain a spin parameter of a black hole. 
For example, in the case of an accretion disk with an inner edge between 
an event horizon and a marginally stable orbit, 
a spin parameter of a black hole is constrained between 0.65 and 0.83 
for an inclination angle of $i=45^\circ$ and a horizontal shift of $0.3$. 
Here, we assume a geometrically thin accretion disk.  
In order to measure a horizontal shift of a black hole shadow, 
which is a good indicator of a spin parameter of a black hole,  
we need to know a position of a rotation axis of a black hole and 
a position of a center of a black hole shadow.  
However, 
it seems to be difficult to determine a position of a rotation axis 
of black holes observationally. 
So, 
although a horizontal shift of a black hole shadow is a good indicator 
of a spin parameter of a black hole, 
this method is not a practical method 
to determine a spin parameter of a black hole by observation of a black hole shadow. 
If we can determine a position of a rotation axis of a black hole by some methods, 
this method is practical for determination of a spin parameter of a black hole. 
In the next section, 
we consider another method to determine a spin parameter of a black hole from 
an image of a black hole shadow. 

\section{Can we measure a spin parameter of a black hole in an accretion disk by 
observation of a black hole shadow?}
\subsection{Quite Similar Black Hole Shadows for Largely Different 
Spin Parameters of Black Holes}


In principle, 
we can determine a spin parameter of a black hole $a/m$, 
an inclination angle $i$ and an innermost radius of an accretion disk 
by fitting a calculated contour 
of a black hole shadow to an observed image of a black hole shadow obtained 
by observational facilities with sufficient spatial resolutions 
and sufficient detection limits even 
in the case that the position of a black hole shadow are not known. 
However, it is practically difficult to determine spin parameters of black hole 
shadows by this fitting method. 
This is because there are black hole shadows 
with a quite similar shape and size for largely different spin parameters 
of black holes. 
In figure 5, we show examples of such black hole shadows with nearly same 
maximum width of a black hole shadow and nearly same ratio of the minimum width 
to the maximum width of a black hole shadow.   
The maximum widths of the black hole shadows are $\sim 6.2$ for black hole 
shadows in panels of (a), (b) and (c), 
and $\sim 7.7$ for black hole shadows in panels of (d), (e) and (f), respectively.  
The flatness ratios are $\sim 0.86$ for black hole shadows in panels of (a), (b) 
and (c), and 
$\sim 0.60$ for black hole shadows in panels of (d), (e) and (f), respectively.  
Here, the adapted parameters are 
(a) $a/m=0$, $i=45^\circ$ and $r_{\rm in}=r_{{\rm h}+}$, 
(b) $a/m=0.5$, $i=45^\circ$ and $r_{\rm in}=1.05r_{{\rm h}+}$, 
(c) $a/m=0.958$, $i=45^\circ$ and $r_{\rm in}=r_{\rm ms}$, 
(d) $a/m=0$, $i=80^\circ$ and $r_{\rm in}=r_{{\rm h}+}$, 
(e) $a/m=0.5$, $i=80^\circ$ and $r_{\rm in}=1.05r_{{\rm h}+}$ and
(f) $a/m=0.955$, $i=80^\circ$ and $r_{\rm in}=r_{\rm ms}$. 
In each panel of figure 5, 
a position of a black hole mass center is plotted by a white dot.   
a rotation axis of an accretion disk is plotted by a gray line. 
In the case of a rotating black hole, this line is also a rotation axis of 
a black hole. 
A horizontal direction in each panel of figure 5 
represents a direction of the maximum width of a black hole shadow. 
As shown in figure 1 and figure 4, 
in the case of a finite spin parameter of a black hole, 
a black hole shadow is non-axisymmetric with respect to a rotation axis of 
a black hole. 
In figure 5, 
extents of shifts of centers of black holes from rotation axes of black holes are 
(b) 0.22, (c) 0.38, (e) 0.50 and (f) 0.95, respectively. 
As mentioned above 
these shifts are nearly proportional to spin parameters of black holes (see figure 4).
In the case of a rotating black hole a direction of a segment of 
the maximum width of a black holes is not perpendicular to a rotation axis of a black hole. 
Thus, as panels of (b), (c), (e) and (f) show, 
rotation axes of black holes are inclined from vertical directions. 
These inclination angles from vertical directions are 
(b) $6.2^\circ$, (c) $12^\circ$, (e) $4.2^\circ$ and (f) $9.2^\circ$, respectively. 
One of the largely different points between black hole shadows with a similar shape and a size 
is a distance between a rotation axis of an accretion disk and a center of a black 
hole shadow. 
If we know a position of a rotation axis of a black hole, 
a spin parameter of a black hole can be well constrained from an image of a black hole shadow. 
But, as mentioned in the previous section, 
it seems difficult to determine a position of a rotation axis of a black hole.  

Here, we newly introduce a bisector axis of a black hole shadow named {\it a shadow axis}. 
We define this shadow axis as a bisector 
perpendicular to a line segment of the maximum width of a black hole shadow. 
In each panel of figure 5, we plot this shadow axis 
by a white line in the case of a rotating black hole.  
In the case of a non-rotating black hole, 
a shadow axis coincides with a rotation axis of a black hole shadow. 
We can easily determine a position of the shadow axis  
by observation of a black hole shadow in an accretion disk. 
In figure 5, 
the minimum intervals between a mass center of a black hole and a shadow axis are 
(b) 0.32, (c) 0.58, (e) 0.59 and (f) 1.1, respectively. 
This minimum interval strongly depends on a spin parameter of a black hole and 
can be measured if a position of a mass center of a black hole is known and a black hole shadow 
is observed. 
So, in the next section, 
we investigate a relation between a spin parameter of a black hole and 
the minimum interval between a mass center of a black hole and a shadow axis.  

\subsection{Minimum Interval between a Black Hole Mass Center and a Shadow Axis} 
In figure 6, we plot 
the minimum interval between a black hole mass center and a shadow axis
for all range of a spin parameter of a black hole. 
Figure 6 is similar to figure 4, 
but there are differences between lines in figure 4 and those in figure 6. 
In figure 6, we plot the minimum intervals for 
three inclination angles ($45^\circ$, $80^\circ$ and $89^\circ$) 
and two kinds of 
accretion disks with $r_{\rm in}=r_{{\rm h}+}$ ({\it dotted lines}) 
and $r_{\rm in}=r_{\rm ms}$ ({\it solid lines}).  
As mentioned in \S 3.1, 
for a non-rotating black hole around an axisymmetric luminous accretion disk 
with respect to a rotation axis of a black hole,  
a black hole shadow is symmetric with respect to a rotation axis of a black hole. 
Thus, in figure 6 for a non-rotating black hole, i.e. $a/m=0$, the minimum interval is null. 
On the other hand, in the case of a rotating black hole, i.e. $a/m>0$, 
the shape and position of the black hole shadow is not symmetric 
with respect to a rotation axis of a black hole. 
So, in such case the minimum interval is finite. 
In figure 6, we can see that 
an extent of the minimum interval is roughly proportional to 
a spin parameter of a black hole for a fixed inclination angle. 
The maximum extent of the minimum interval is about $1.5$ 
in the case of the maximum rotation, the maximum inclination angle and 
$r_{\rm in}=r_{{\rm h}+}$. 
The minimum interval for the case of $r_{\rm in}=r_{{\rm h}+}$ 
is larger than the minimum interval for the case of $r_{\rm in}=r_{\rm ms}$ 
for same inclination angle and same spin parameter of a black hole. 
This is because effects of frame dragging due to black hole rotation 
are more significant in a nearer region of a black hole, 
as mentioned in the previous section. 
If we can measure such minimum interval, 
a spin parameter of a black hole can be well constrained. 
But if not, quite subtle fitting of a contour of a black hole shadow 
to an observed image of a black hole shadow 
is required in order to determine a spin parameter of a black hole 
and other physical parameters from an observed image of a black hole shadow. 
For example, for an accretion disk with an inner edge between 
an event horizon and a marginally stable orbit  
with an inclination angle of $i=45^\circ$ and a horizontal shift of a black hole shadow 
to be $0.5$, from figure 6 
a spin parameter of a black hole is constrained between 0.75 and 0.89. 
Thus, 
{\it 
in order to measure a spin parameter of a black hole 
it is crucially important to determine 
a position of black hole mass center in a region of a black hole shadow. 
}
Required positional accuracy to determine a spin parameter of a black hole 
is lower than about $1.5$ (in units of $r_g$). 
This is the case of the maximum shift of black hole shadow center 
from a rotation axis. 
In the last section, we discuss determination methods of a mass center of a black hole. 


Note that 
in order to determine a spin parameter of a black hole 
from an image of a black hole shadow 
we do not always need to determine a position of a black hole mass center. 
For some cases, 
we can determine a spin parameter of a black hole by fitting a calculated contour 
of a black hole shadow to an observed image of a black hole shadow. 
One of the examples of such cases is  
a black hole shadow with tips which can be seen when a relatively 
large inner edge of an accretion disk as mentioned in \S 3.1. 
In bottom panels of figure 1, 
in the case of $r_{\rm in}=r_{\rm ms}$ there are tips in contours of black hole 
shadows for $a/m=0$ and $0.5$. 
When $a/m=0$ there are two tips, when $a/m=0.5$ there is a single tip and 
when $a/m=0.998$ there is no tips. 
So, if there are one or two tips in contours of black hole shadows, 
fitting of a calculated contour of a black hole shadow to an observed black hole 
shadow is more easy than the cases with no tips in a contour of a black hole shadow.   
But, the black hole shadow with tips is a quite limited case. 

\section{Discussion and Conclusions}
\subsection{How is the Position of the Center of Mass of a Black Hole Determined?}
As mentioned in the previous section, 
in order to measure a spin parameter of a black hole from an observed image of 
a black hole shadow 
it is important to determine a position of a black hole mass center. 
In the case of Sgr A*, stellar motions around black holes are actually observed 
(e.g. Genzel et al. 2003). 
From these stellar motions, in principle we can determine the position of 
a mass center of a black hole. 
Astrometry satellite missions with positional accuracy of 10 $\mu$ arc second 
which will give a plenty of data of stellar motions around Sgr A* are planned: 
GAIA for optical observation (e.g. Belokurov \& Evans 2002 and references therein)  
and JASMINE (Japan Astrometry Satellite Mission for INfrared Exploration) (Gouda 2002) 
for infrared observation in which effects of interstellar absorption and scattering 
are not very significant compared to those of optical observation. 

There is a simple additional method to determine a black hole mass center. 
Distribution of brightness of the radiation from the accreting material 
around the black hole are influenced by the Doppler and gravitational 
boosting effects. 
For example, Luminet (1979) provides detailed derivations and discussions 
of the brightness distribution over the image of the accretion disk. 
On the other hand, an angular momentum, $\omega$, of the frame dragging due to 
a black hole rotation is described as, 
\begin{eqnarray}
\omega&=&-\frac{g_{t{\phi}}}{g_{\phi\phi}}\\
	&=&\frac{2mra}{(r^2+a^2)^2-a^2\Delta\sin^2\theta}. 
\end{eqnarray}
In the equatorial plane, this angular momentum is expanded as 
\begin{equation}
\omega=\frac{2ma}{r^3}-\frac{2ma^3}{r^4}+O\left(\frac{1}{r^5}\right). 
\end{equation}
Because the frame dragging effect decreases in proportion to $a/r^3$ as shown here  
for a sufficiently large radial distance, $r$, 
from a mass center of a black hole,
and the gravitational potential decreases only proportional to $1/r$, 
the frame dragging effects on the propagation of photons become 
negligible at a sufficient distance from the black hole. The image at 
these large radii is formed only by the influence of Doppler effect from 
the moving gas and gravitational redshift in a spherically symmetric 
potential. This image should be clearly visible, since the shadow of
the black hole is assumed to be clearly visible. If the motions and 
density distribution of the gas are axisymmetric, than the image at 
large radii would be the same as produced by a non-rotating black hole. 
One can then use calculations by Luminet (1979) to fit the observed
brightness distribution and determine the center of mass and the 
position of the rotation axis of the black hole as the parameters of 
this fit. Then, it makes sense to use both figure 4 and figure 6 
to constrain the spin parameter of the black hole. 

\subsection{Observational Feasibility of Black Hole Shadows}
Although in the present study we investigate black hole shadows 
in accretion disks from mainly academic interests as several past 
studies about black hole shadows in accretion disks 
(e.g. Luminet 1979, Fukue \& Yokoyama 1988, Fukue 2003), 
observational feasibilities of black hole shadows must be considered. 
Falcke, Agol \& Melia (2000) shows that 
a black hole shadow in Sgr A* can be observed at sub-millimeter wavelengths 
at which a scattering radius due to interstellar 
medium is sufficiently small in order to observe a black hole shadow. 
Also, in nearby elliptical galaxy M87 there is a slightly smaller 
black hole shadow than that in Sgr A*. 
Both of observed spectrum of SgrA* and M87 will be well reproduced by 
advection dominated accretion flows (ADAFs) or recently 
radiatively inefficient accretion flows (RIAFs) for Sgr A*. 
In such accretion models, matter densities are quite low and 
it is expected that optically thick accretion disks are not formed. 
So, black hole shadows in Sgr A* and M87 seems to be seen as 
black hole shadows not in accretion disks but 
in optically thin medium as Falcke, Melia \& Agol (2000) have calculated. 
In the case of M87, Junor, Viretta \& Livio (1999) report 
observations at 43 GHz of inner regions of M87 and shows 
a remarkably broad jet near the center and strong collimation of the 
jets occurring at $\sim30-100$ Schwarzschild radii from a black hole. 
These results are consistent with the hypothesis that 
jets are formed by an accretion disk around the central black hole. 
In the case of M87 
we can not completely deny the observational feasibility of a black hole shadows 
in an accretion disk by next-generation radio interferometers. 
In such cases, the present study about black hole shadows in accretion disks 
is applicable to the black hole shadow in M87. 
The central region of M87 is one of the observational targets of 
the planned next-generation VLBI Space Observatory Program ({\it VSOP-2}) 
(Hirabayashi et al. 2001) with extremely high resolution. 
Theoretically, 
the black hole shadow in the center of M87 need more studies 
which include radiative transfer calculations with Synchrotron source functions of 
accretion disks and jets. 

\subsection{Conclusions}
In the present study, 
we make a qualitative analysis and 
a quantitative analysis of the shape and position of the black hole shadow 
casted by a rotating black hole on the optically thick accretion disk and its dependence 
on the angular momentum of the black hole. 
We consider two types of innermost radii, an event horizon and a marginally stable orbit, 
of an accretion disk. 
There are black hole shadows with quite similar sizes and shapes for largely 
different spin parameters of black holes and same masses of black holes. 
In the present study, we newly introduce a bisector axis of a black hole shadow named a shadow axis. 
We define this shadow axis as a bisector 
perpendicular to a line segment of a maximum width of a black hole shadow. 
We can determine a position of shadow axis of a black hole shadow by observation 
of a black hole shadow. 
The minimum interval between a black hole mass center and a shadow axis is important. 
This is because the minimum interval is quite roughly proportional to 
a spin parameter of a black hole for a fixed inclination angle. 
The maximum extent of the minimum interval is about $1.5$ 
in the case of the maximum rotation, the maximum inclination angle and 
$r_{\rm in}=r_{{\rm h}+}$. 
If we can measure such minimum interval, i.e. if we can determine 
positions of a black hole mass center and a shadow axis,  
a spin parameter of a black hole can be well constrained. 
Determination methods of a mass center of a black hole are discussed in the previous 
subsection. 

Lastly, we adress the assumptions used in the present study.  
In the present study we only consider a geometrically thin accretion disk and  
calculate direct geodesics when calculating black hole shadows. 
An accretion disk with finite thickness may cause self-eclipsing effects of an accretion 
disk for large inclination angle and a black hole shadow may be deformed. 
So, this self-eclipse effect could be important for high inclination angles 
and need study in future. 
There are also geodesics which come from an underside 
of accretion disks. These photons also changes shapes of black hole shadows. 
This effect also need study in future. 

\acknowledgments
The author is grateful to 
the referee, V. I. Pariev, for helpful comments on discussion, 
K. Ohsuga, K. Watarai and Y. Kato for their useful discussions, 
H. Hirabayashi, H. Kamaya and S. Mineshige for their useful comments 
and continuous encouragement.  
This work was supported in part by the Grant-in Aid of the Ministry of Education, 
Culture, Sports, and Science and Technology.



\begin{figure}
\epsscale{0.85}
\plotone{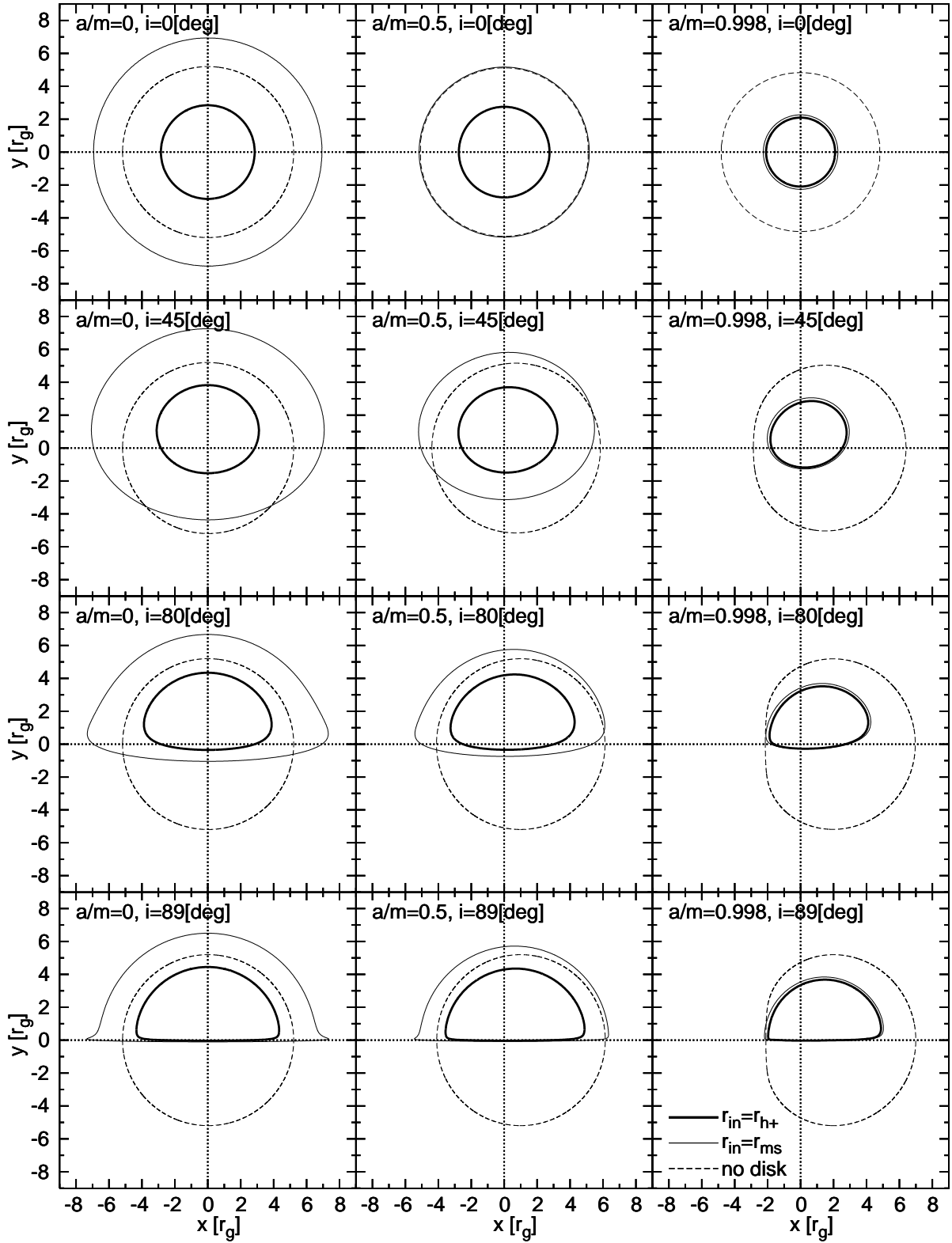}
\caption{
Examples of contours of black hole shadows 
in geometrically thin optically thick 
accretion disks with innermost radii, $r_{\rm in}$, of
$r_{\rm in}=r_{{\rm h}+}$ ({\it thick solid lines}) and 
$r_{\rm in}=r_{\rm ms}$ ({\it thin solid lines}) and in optically thin 
emitting medium ({\it dashed lines}). 
Lines of $x=0$ and $y=0$ are plotted by dotted lines. 
A projected position of a black hole is $(x,~y)=(0,~0)$. 
Projected lines of rotation axes of accretion disks or 
those of black holes are $x=0$. 
In the cases of a finite spin parameter of a black hole and a finite inclination
angle, we can see an axisymmetric shape and position of a black hole shadow 
with respect to a rotation axis of an accretion disk. 
\label{shape}
}
\end{figure}

\begin{figure}
\epsscale{1.0}
\plotone{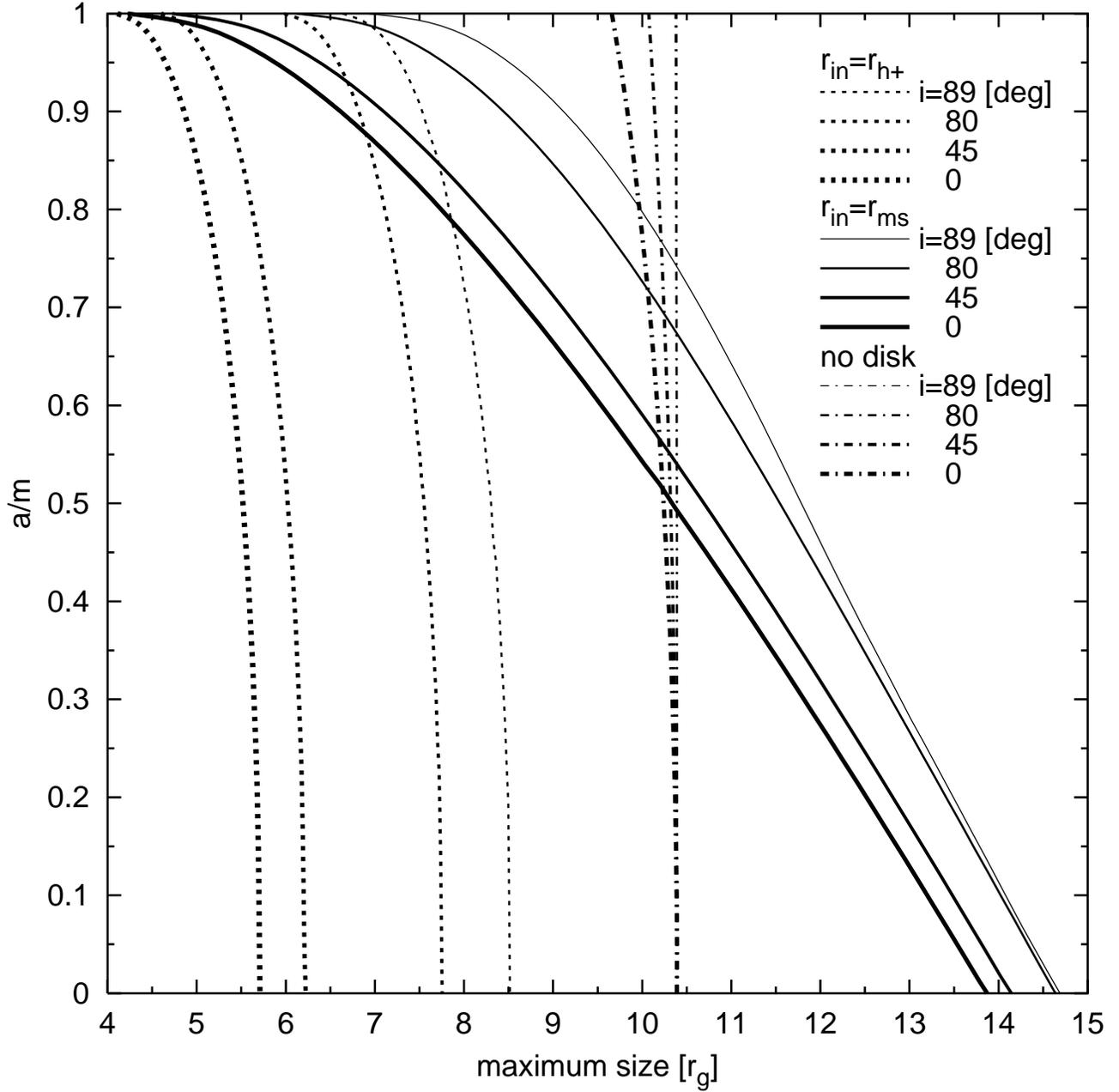}
\caption{
Relations between a spin parameter of a black hole and  
the maximum size of a black hole shadow for four values of inclination angles 
($0^\circ$, $45^\circ$, $80^\circ$ and $89^\circ$) and three kinds of 
surrounding mediums; 
accretion disks 
with innermost radius of $r_{\rm in}=r_{{\rm h}+}$ ({\it dotted lines})
and those with $r_{\rm in}=r_{\rm ms}$ ({\it solid lines}), 
and in optically thin emitting medium ({\rm dash-dotted lines}). 
\label{maxBH}
}
\end{figure}

\begin{figure}
\epsscale{1.0}
\plotone{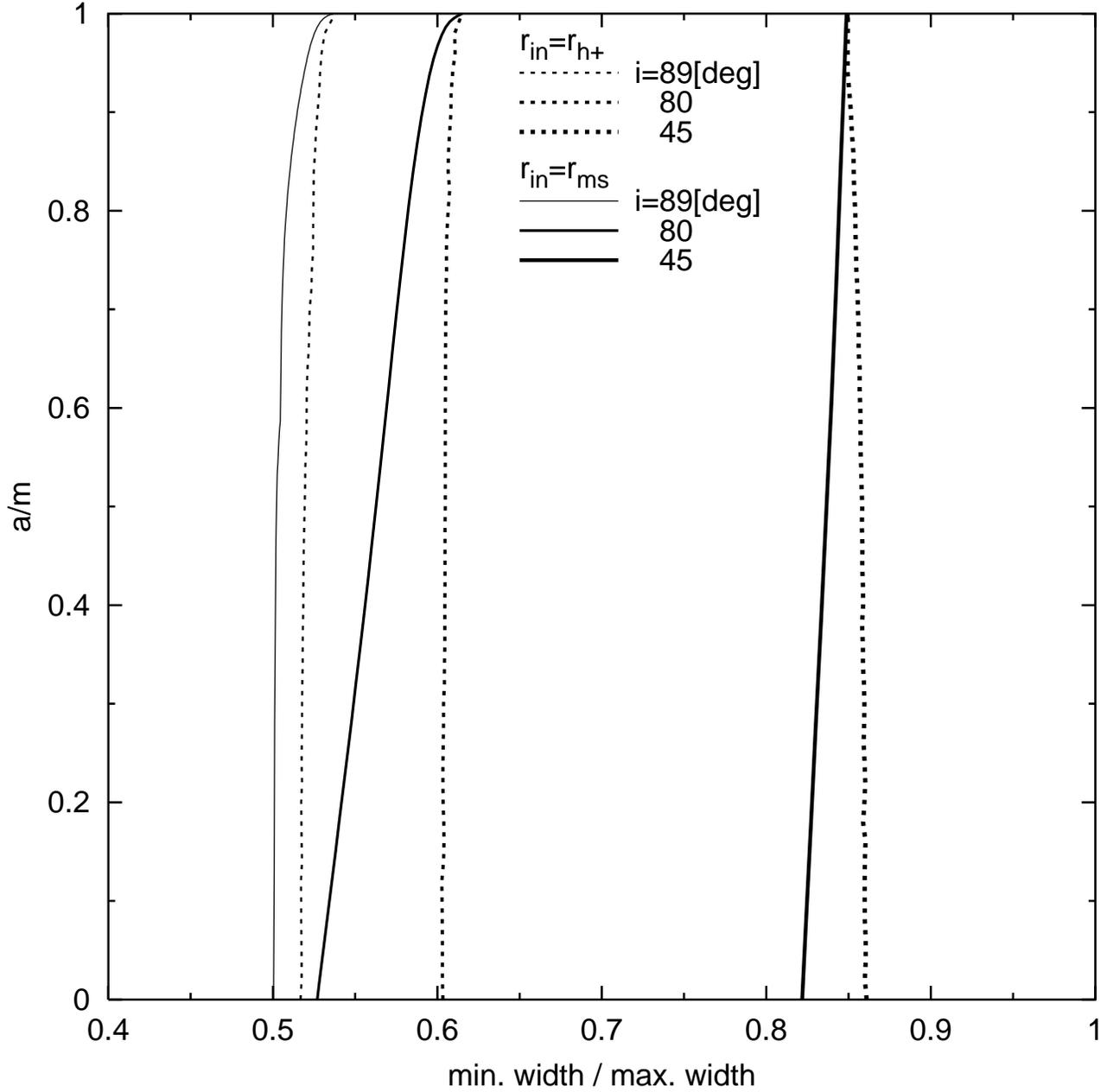}
\caption{
Ratios of the minimum size of a black hole shadow to the maximum size of 
a black hole shadow for three inclination angles ($45^\circ$, $80^\circ$ 
and $89^\circ$) and two kinds of 
accretion disks with $r_{\rm in}=r_{{\rm h}+}$ ({\it dotted lines}) 
and $r_{\rm in}=r_{\rm ms}$ ({\it solid lines}).  
In the case of an inclination angle of $i=0^\circ$, this ratio is unity. 
The ratio strongly depends on an inclination angle between 
a rotation axis and a direction of an observer, 
and weakly depends on a spin parameter of a black hole. 
\label{ratios}
}
\end{figure}


\begin{figure}
\epsscale{1.0}
\plotone{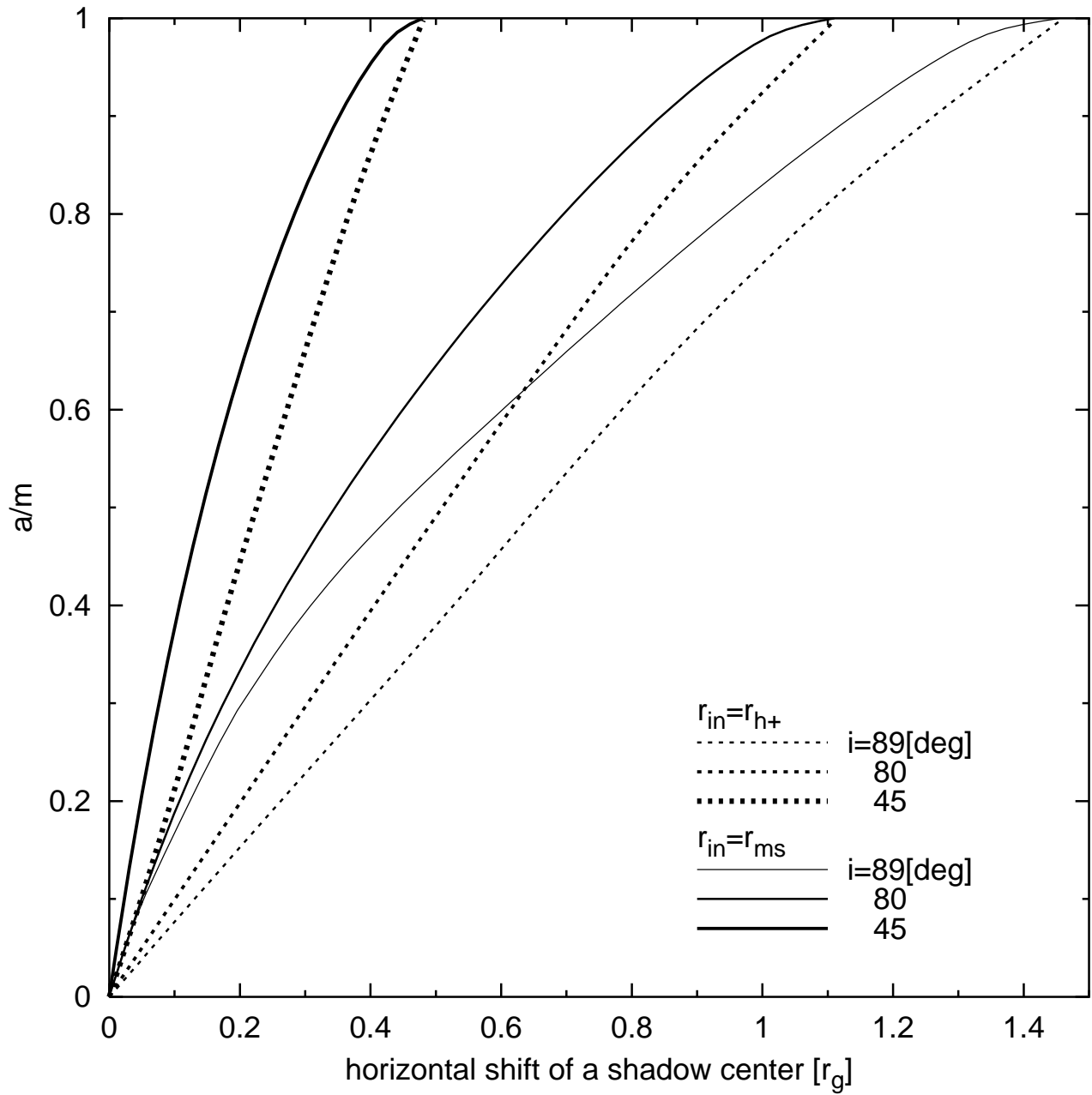}
\caption{
Shifts of centers of black hole shadows in the direction of $x$ axes 
from rotation axes 
for three inclination angles ($45^\circ$, $80^\circ$ and $89^\circ$) 
and two kinds of 
accretion disks with $r_{\rm in}=r_{{\rm h}+}$ ({\it dotted lines}) 
and $r_{\rm in}=r_{\rm ms}$ ({\it solid lines}).  
As this figure shows, 
the shift is nearly proportional to a spin parameter of a black hole.
In the case of an inclination angle of $i=0^\circ$, 
an extent of this shift is null. 
If we can know an extent of a horizontal shift of a black hole shadow from 
a rotation axis of a black hole and an inclination angle, 
a spin parameter of a black hole is well constrained. 
\label{shifts}
}
\end{figure}

\begin{figure}
\epsscale{0.9}
\plotone{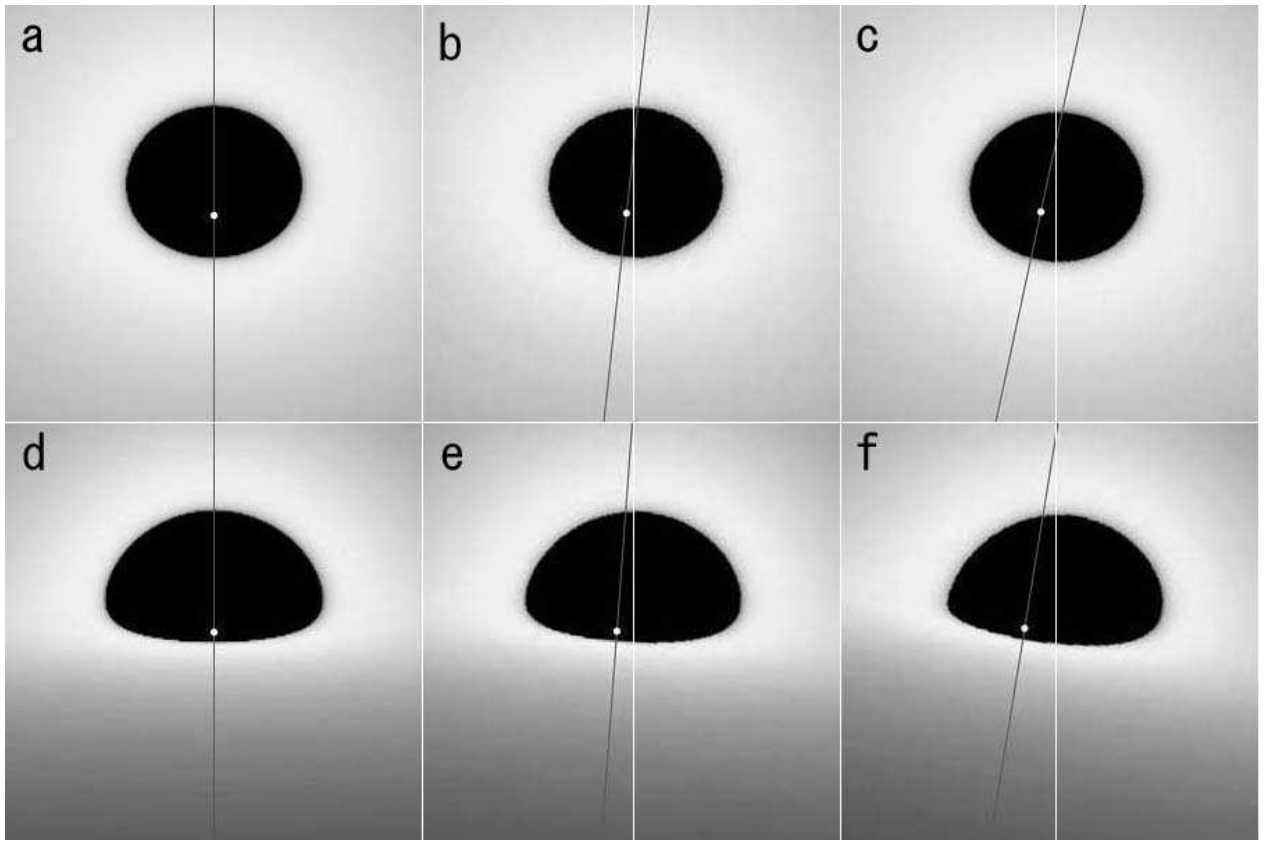}
\caption{
Examples of black hole shadows with nearly same maximum width and flatness 
for largely different spin parameters of black holes. 
Refer to the main text for the explanation. 
\label{similarBH}}
\end{figure}

\begin{figure}
\epsscale{1.0}
\plotone{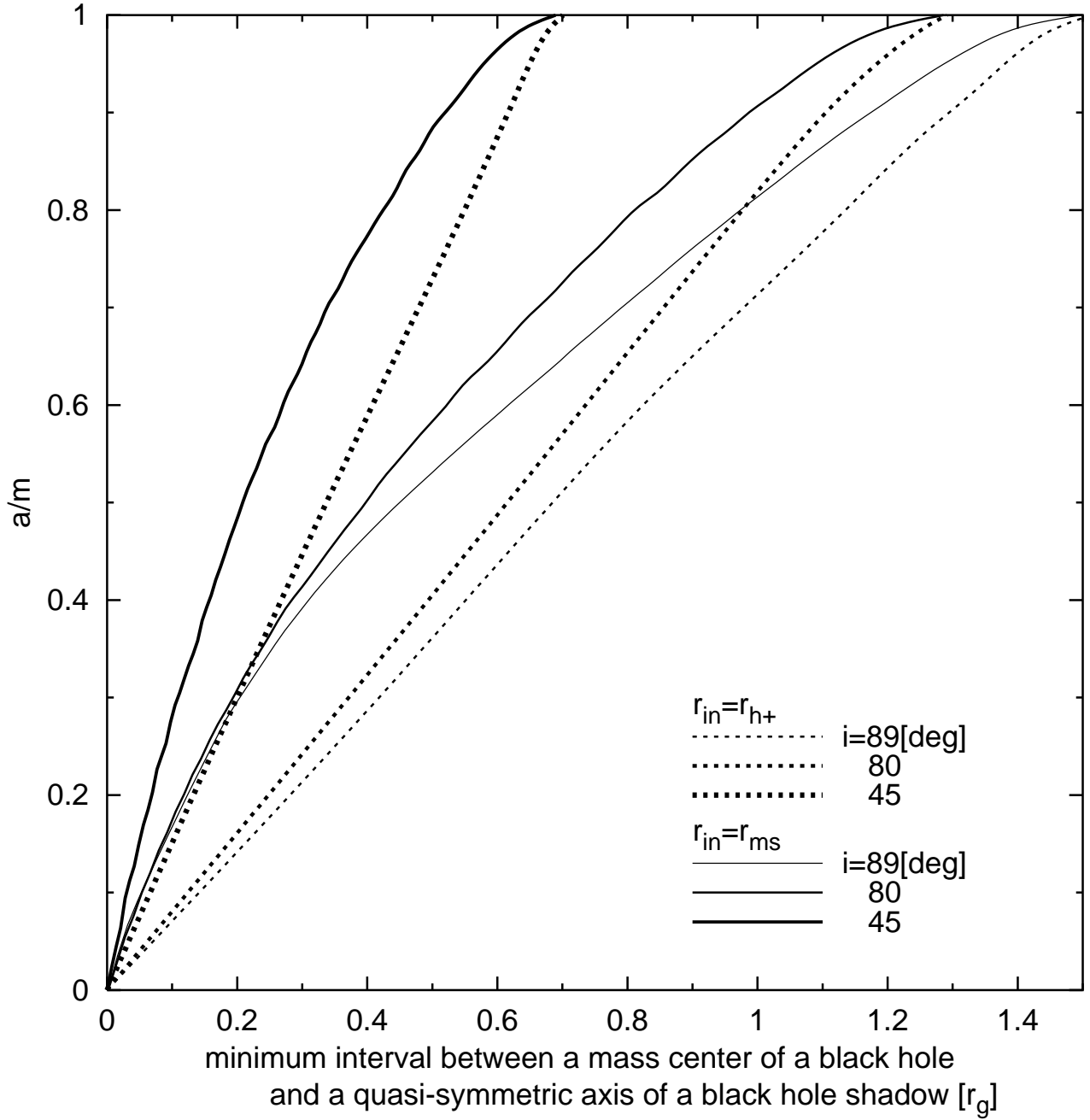}
\caption{
Relations between a spin parameter of a black hole and 
the minimum interval between a mass center of a black hole and a shadow axis 
for three inclination angles ($45^\circ$, $80^\circ$ and $89^\circ$) 
and two kinds of 
accretion disks with $r_{\rm in}=r_{{\rm h}+}$ ({\it dotted lines}) 
and $r_{\rm in}=r_{\rm ms}$ ({\it solid lines}).  
\label{shifts2}
}
\end{figure}

\end{document}